\documentclass{elsart}
\usepackage{epsfig}
\usepackage{epsf}
\usepackage{amssymb}
\begin{document}
\begin{frontmatter}
\title{Effective critical behaviour of diluted Heisenberg-like
magnets}
\author[icmp]{M. Dudka}
\ead{maxdudka@icmp.lviv.ua}
\author[kepler]{R. Folk}
\ead{folk@tphys.uni-linz.ac.at}
\ead[url]{http://www.uni-linz.ac.at/fak/TNF/theophys/group/folk.html}
\author[icmp,franko]{Yu. Holovatch}
\ead{hol@icmp.lviv.ua}
\ead[url]{http://ph.icmp.lviv.ua/$^{\sim}$hol}
\author[franko]{D. Ivaneiko}
\ead{dmytro$\_$ivaneiko@yahoo.com}
\address[icmp]{Institute for Condensed Matter Physics
of the National Academy of Sciences of Ukraine,
UA-79011 Lviv, Ukraine}
\address[kepler]{Institut f\"ur Theoretische Physik,
Johannes Kepler Universit\"at Linz,
A-4040 Linz, Austria}
\address[franko]{Ivan Franko National University of Lviv,
Chair for Theoretical Physics,
UA-79005 Lviv, Ukraine}
\begin{abstract}
In agreement with the Harris criterion, {\em asymptotic}
critical exponents of three-dimensional (3d) Heisenberg-like
magnets are not influenced by weak quenched dilution of
non-magnetic component. However, often in the experimental studies of
corresponding systems con\-cen\-tration- and
tem\-pe\-ra\-tu\-re-dependent
exponents are found with  values differing from those of the 3d
Heisenberg model.

In our study, we use the field--theoretical renormalization group
approach to explain this observation and to calculate the
{\em effective} critical exponents of weakly diluted quenched
Heisenberg-like magnet. Being non-universal, these exponents change
with  distance to the critical point ${\rm T_c}$ as observed
experimentally.  In the asymptotic limit (at ${\rm T_c}$) they
equal to the critical exponents of the pure 3d Heisenberg magnet as
predicted by the Harris criterion.
\end{abstract}

\end{frontmatter}

%%%%%%%%%%%%%%%%%%%%%%%%%%%%%%%%%%%%%%%%%%%%%%%%%%%
%SECTION I
%%%%%%%%%%%%%%%%%%%%%%%%%%%%%%%%%%%%%%%%%%%%%%%%%%%
\section{Introduction}
\label{I}

Relevance of structural disorder for the critical behaviour remains
to be an important problem of modern condensed matter physics.
Even a weak disorder may change drastically the behaviour near the
critical point and in this respect may be related to the global
characteristics of a physical system, such as the space dimension,
order parameter symmetry and the origin of interparticle interaction.
In this paper, we are going to discuss some peculiarities of a
paramagnetic-ferromagnetic phase transition in magnets, where the
randomness of structure has the form of substitutional random-site
or random-bond quenched disorder. Solid solutions of magnets with
small concentration of non-magnetic component as well as amorphous
magnets with large relaxation times may serve as an example of
such systems.

Intuitively, it is clear that  for a weak enough disorder
the ferromagnetic phase persists in
such systems. Obviously, intuition fails
to predict whether the critical exponents characterizing phase
transition into ferromagnetic state will differ in a disordered
system and in a ``pure" one. The answer here is given by the Harris
criterion \cite{Harris74} which states that the critical exponents
of the disordered system are changed only if the heat capacity
critical exponent of a pure system is positive,
otherwise the critical exponents of a disordered system coincide
with those of a pure one. Returning to $d=3$ dimensional magnets
with $O(m)$ symmetric spontaneous magnetization  one is lead to the
conclusion, that here only the critical exponents of uniaxial magnets
described by the $d=3$ Ising model ($m=1$) are the subject of
influence by weak quenched disorder. Indeed, the heat capacity
diverges $\alpha=0.109\pm0.004>0$ \cite{Guida98} for $m=1$, whereas
it does not diverge for the easy-plane and Heisenberg-like magnets:
$\alpha=-0.011\pm0.004$ and $\alpha=-0.122\pm0.010$ for $m=2$ and
$m=3$, respectively \cite{Guida98}.

Note however that the Harris criterion tells about the scaling
behaviour {\em at} the critical point $T_c$. In other words it
predicts (possible) changes in the {\em asymptotic} values of
the critical exponents defined at $T_c$. In real situations one
often deals with the {\em effective} critical exponents governing
scaling when $T_c$ still is not reached \cite{effective}. These are
non-universal. As far as in our study of particular interest
will be the isothermal magnetic susceptibility $\chi_T$ let us define the
corresponding effective exponent by \cite{effective}:
\begin{equation}\label{1}
\gamma_{\rm eff}(\tau)=-\frac{{\rm d}\ln\chi(\tau)}{{\rm d}\ln\tau},
\hspace{2em}\mbox{with} \hspace{2em} \tau=|T-T_c|/T_c.
\end{equation}
In the limit $T\to T_c$ the effective exponent coincides with the
asymptotic one $\gamma_{\rm eff}=\gamma$.

Already in the first experimental studies
of weakly diluted uniaxial (Ising-like) $d=3$ random magnets \cite{note1} the
asymptotic values of critical exponents were found. For the solid solutions,
the exponents do not depend on the concentration of non-magnetic component
and belong to the new universality class \cite{review} as predicted
by the Harris criterion.
We do not know analogous experiments where an influence of
disorder on criticality of easy-plane magnets was examined. However its
irrelevance  was experimentally proven  \cite{helium} for the superfluid phase
transition in ${\rm He^4}$ which belongs to the same $O(2)$ universality class
as  the ferromagnetic phase transition in easy-plane magnets.

\begin{figure}[htbp]
\centerline{
\epsfxsize=52mm\epsfysize=30mm\epsfbox{f2_22.eps}
\hspace{3.2cm}
\epsfxsize=48mm\epsfysize=30mm\epsfbox{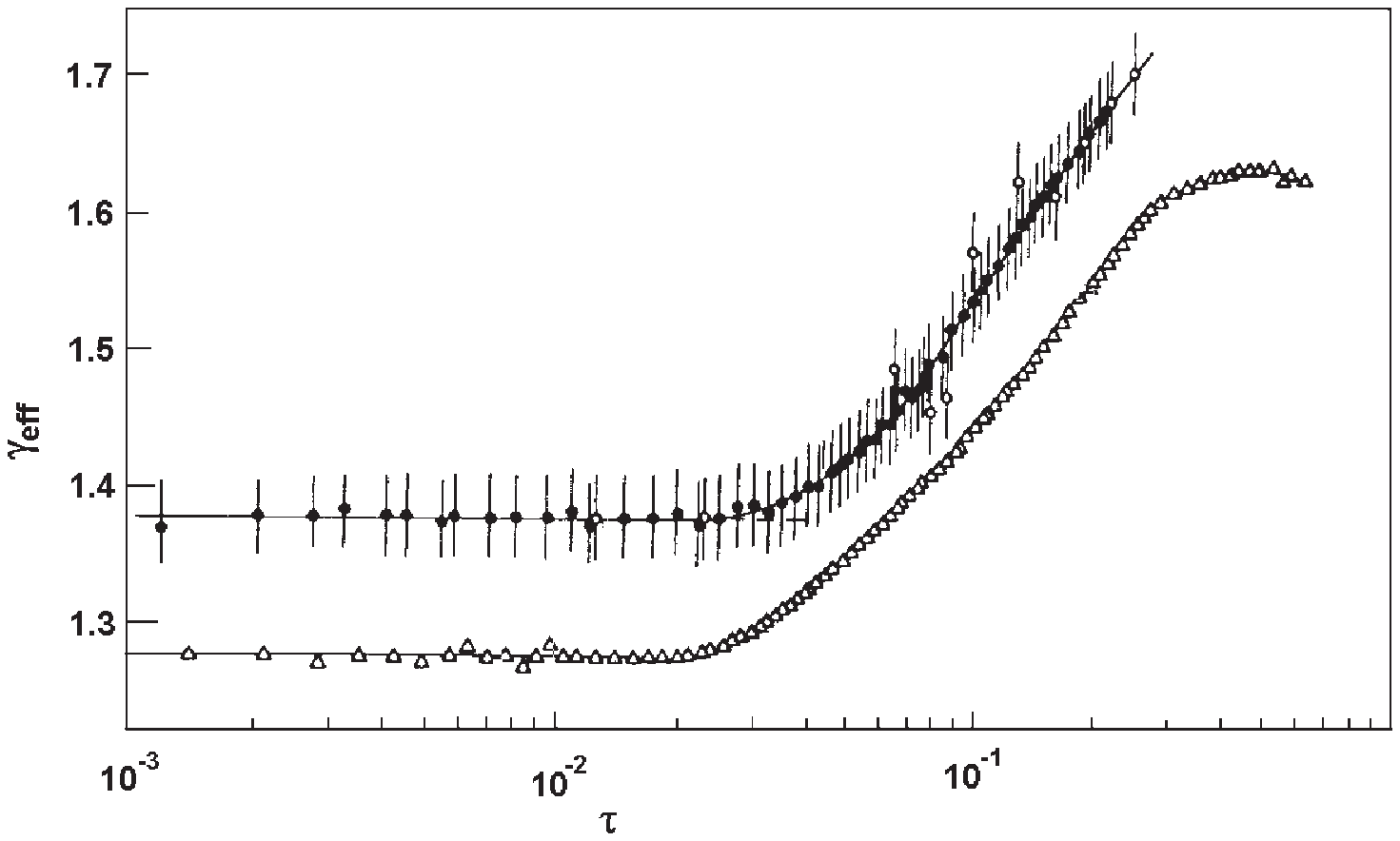}}
{\bf a.}\hspace{7.5cm}{\bf b.}
\vspace{0.5cm}\\
\centerline{
\epsfxsize=50mm\epsfysize=30mm\epsfbox{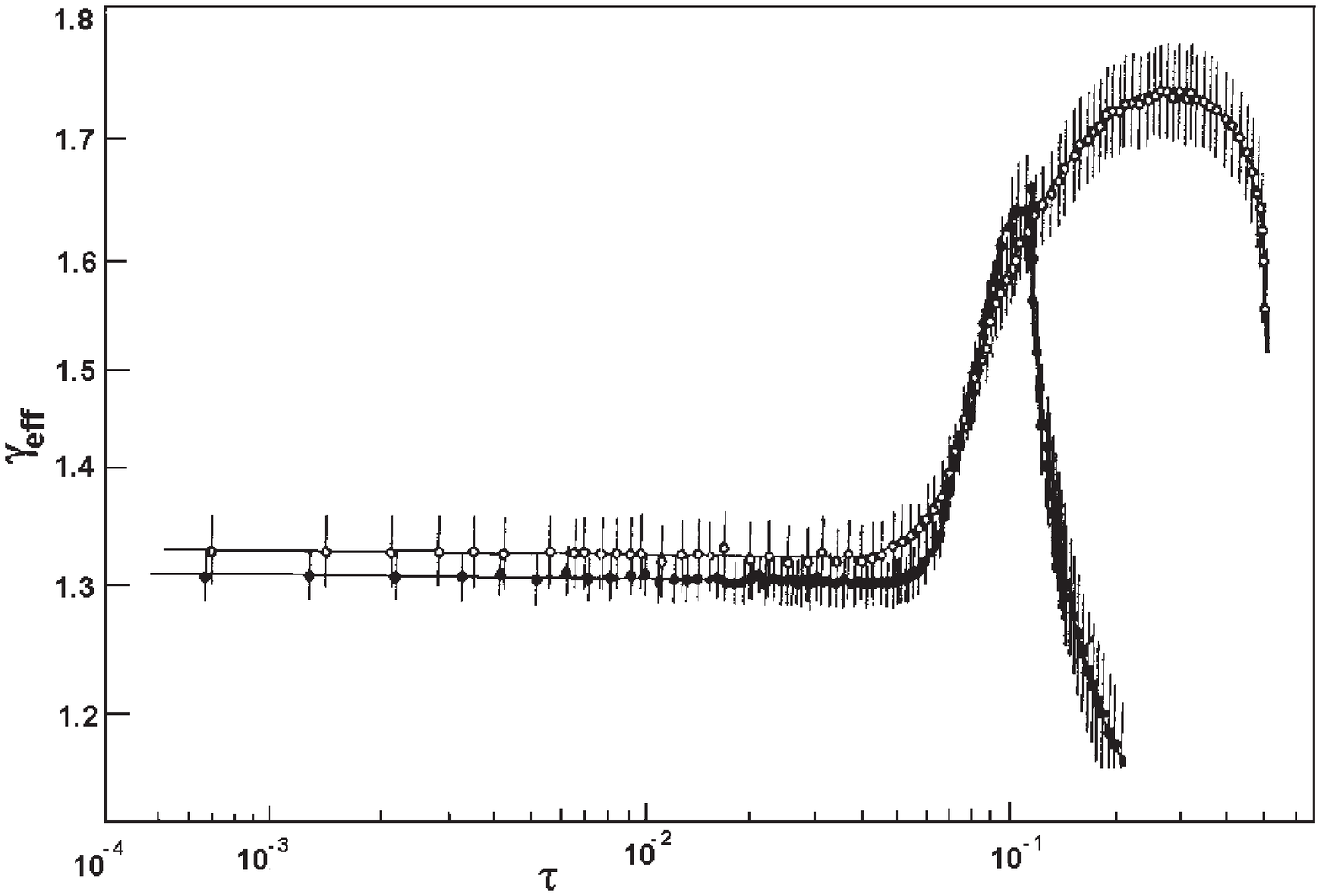}
\hspace{3.2cm}
\epsfxsize=48mm\epsfysize=30mm\epsfbox{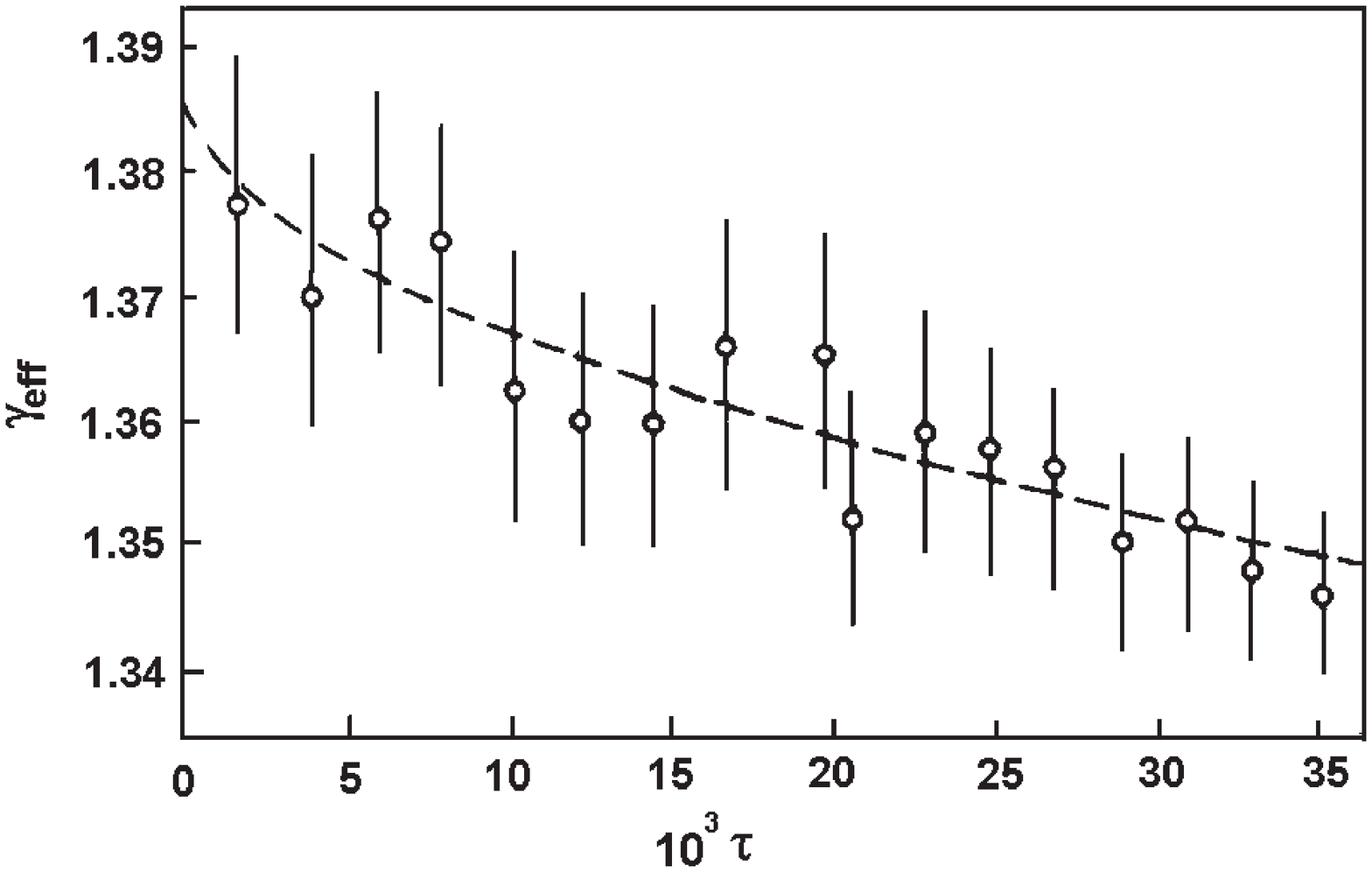}}
\hspace{3cm}{\bf c.}\hspace{7.5cm}{\bf d.}
\vspace{0.9cm}\\
\centerline{
\epsfxsize=52mm\epsfysize=30mm\epsfbox{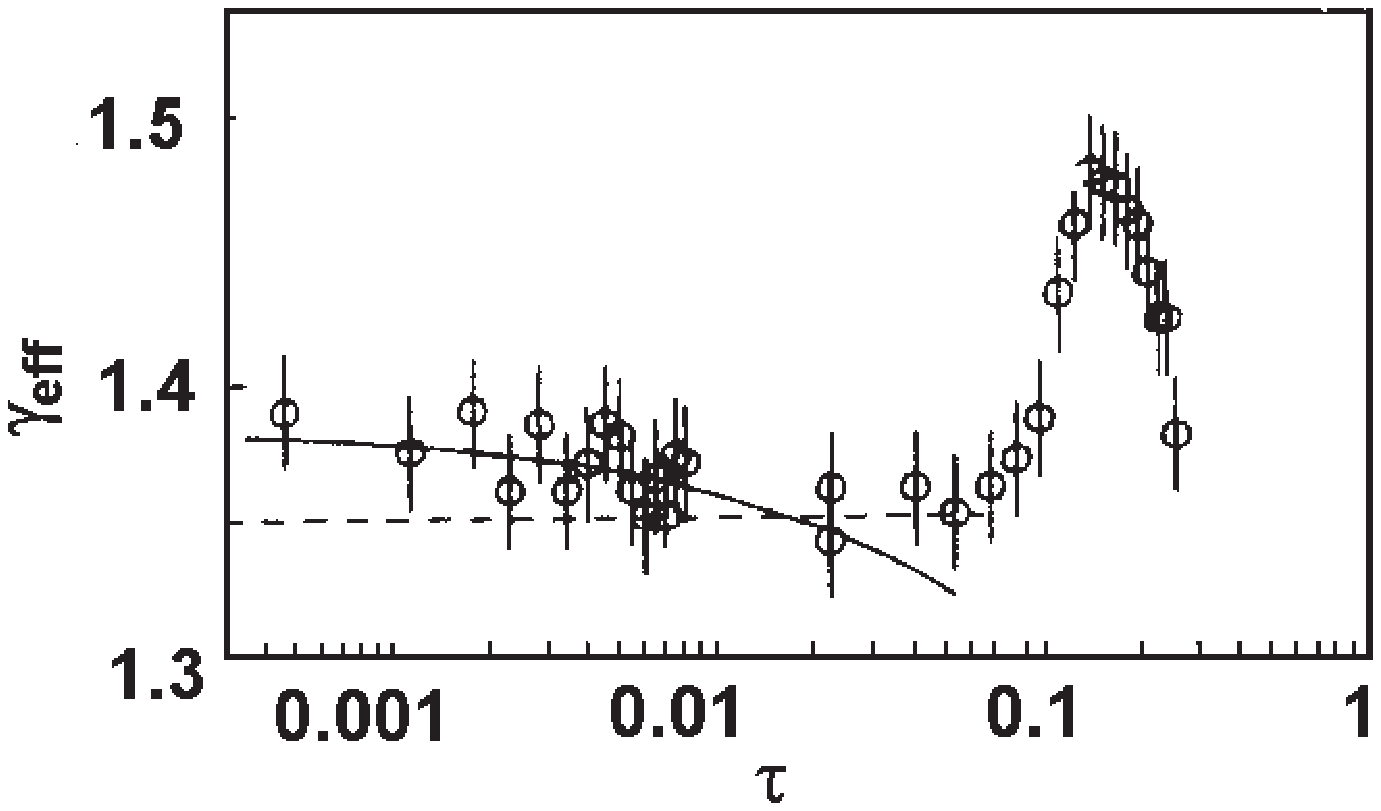}
\hspace{2.8cm}
\epsfxsize=51mm\epsfysize=30mm\epsfbox{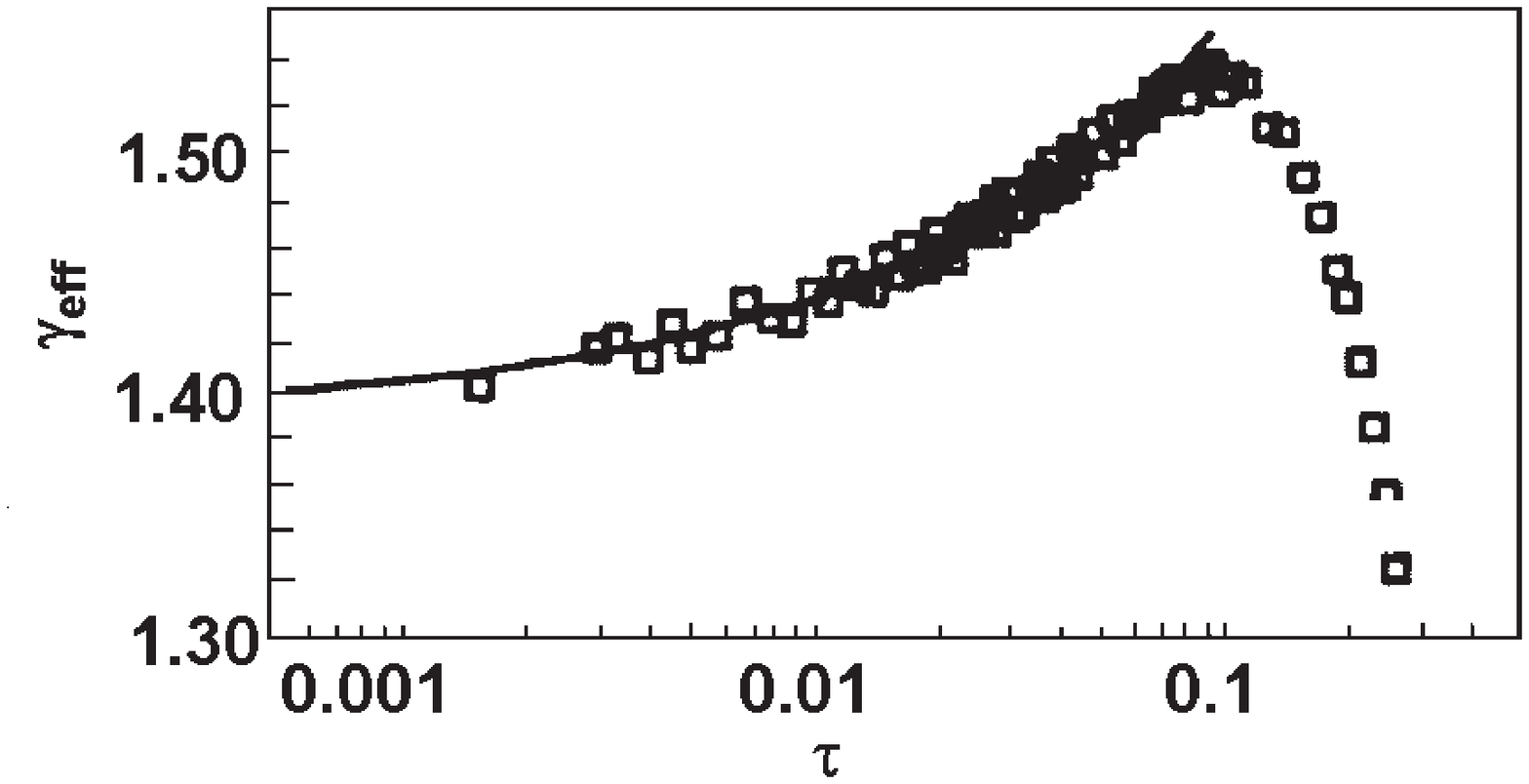}}
\hspace{3cm}{\bf e.}\hspace{7.5cm}{\bf f.}
\vspace{0.5cm}\\
\centerline{
\epsfxsize=52mm\epsfysize=30mm\epsfbox{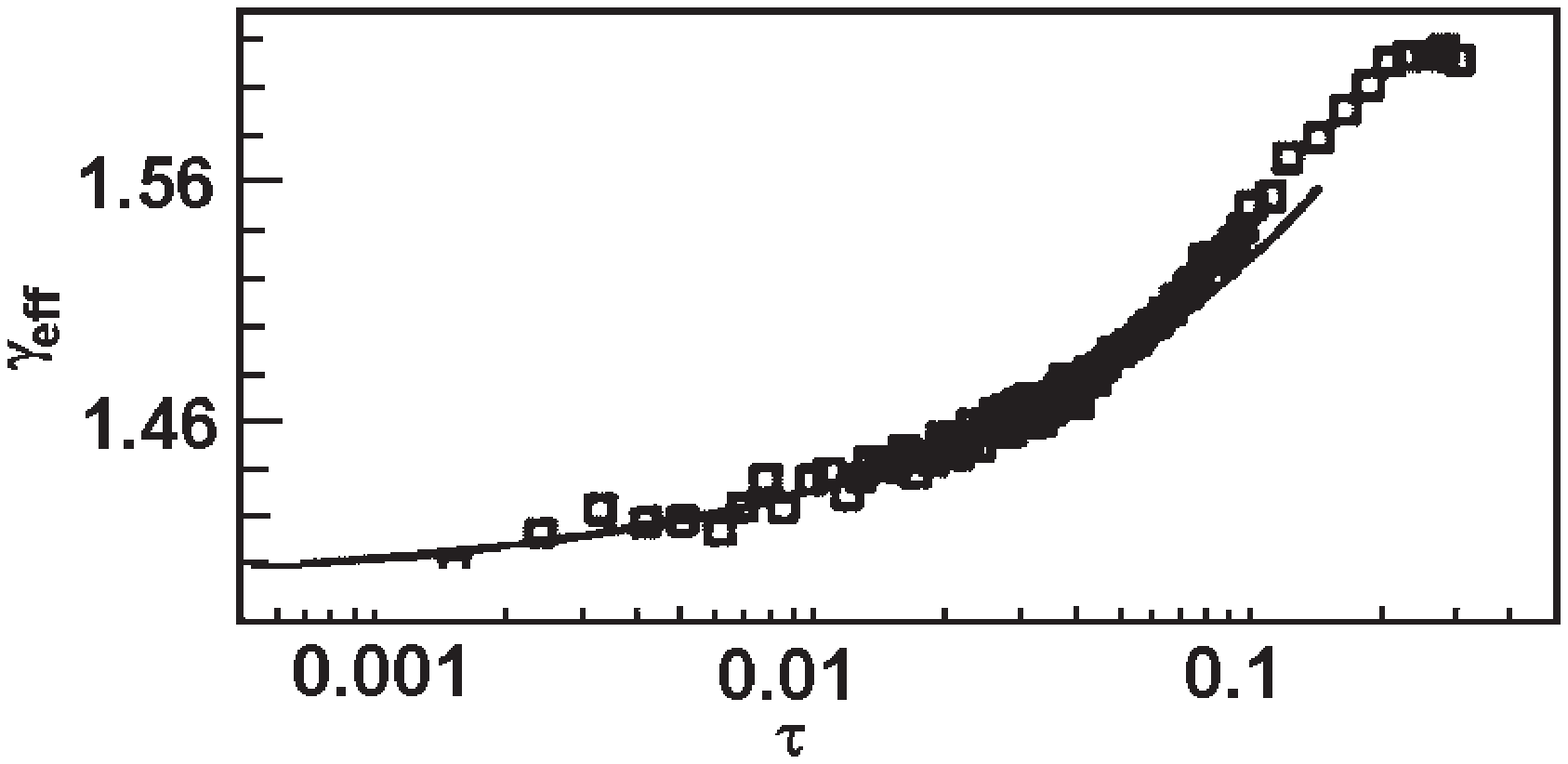}
\hspace{3.2cm}
\epsfxsize=46mm\epsfysize=30mm\epsfbox{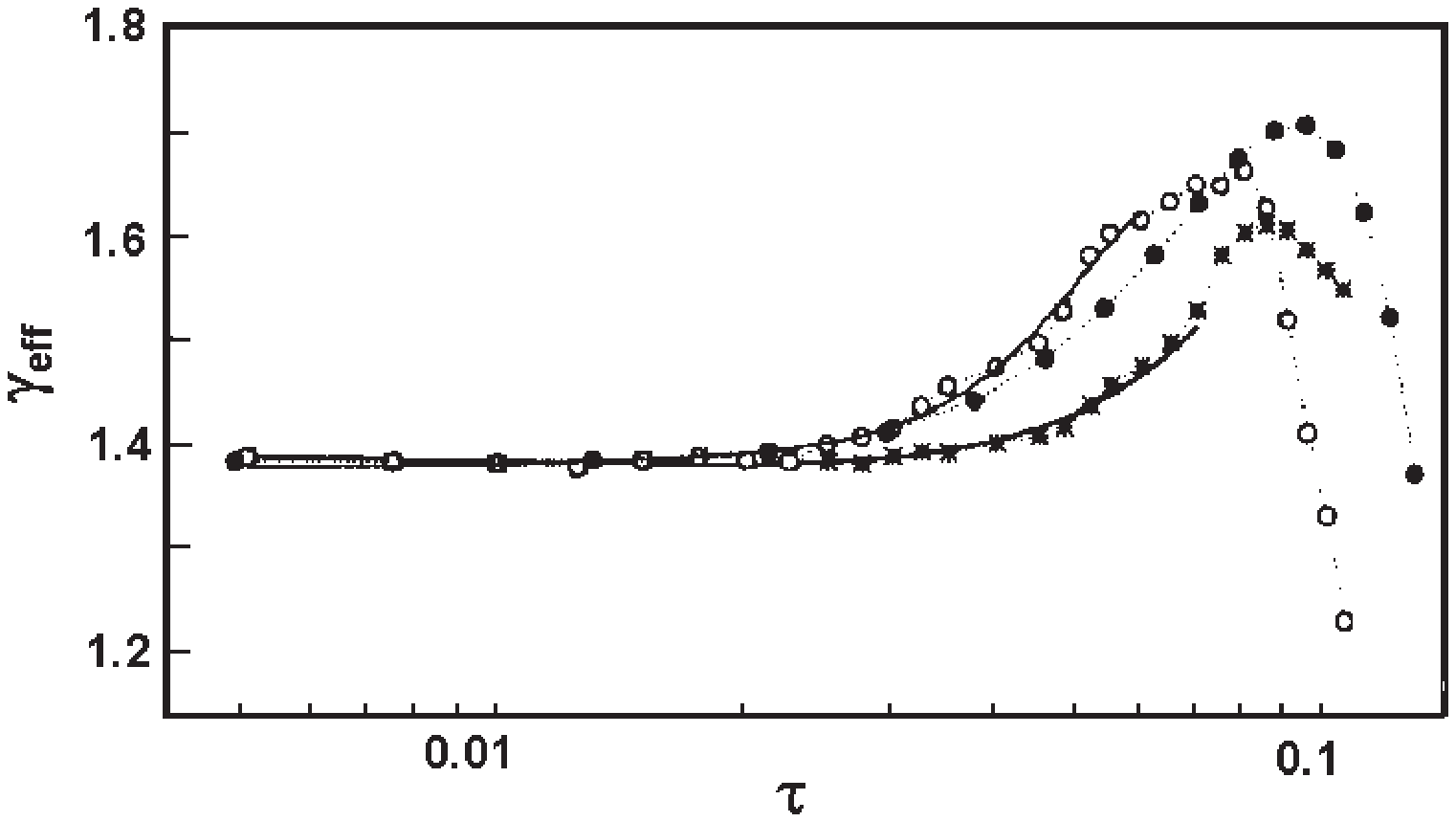}}
{\bf g.}\hspace{7.5cm}{\bf h.}
\vspace{0.5cm}
\caption{\label{fig1} Experimentally measured  isothermal magnetic susceptibility
effective critical exponent $\gamma_{\rm eff}$ for disordered Heisenberg-like
magnets ($\tau=(T-T_c)/T_c$).
{\bf a.}:
${\rm Fe_{20}Ni_{56}B_{24}}$ (F\"ahnle {\em et al.}, 1983 \cite{Fahnle83});
{\bf b.}:
${\rm Fe_{32}Ni_{36}Cr_{14}P_{12}B_6}$ (Kaul, 1985 \cite{Kaul85});
{\bf c.}:
${\rm Fe_{20}Ni_{60}P_{14}B_6}$,
${\rm Fe_{40}Ni_{40}P_{14}B_6}$ (Kaul, 1985 \cite{Kaul85});
{\bf d.}:
${\rm Fe_{10}Ni_{70}B_{19}Si_1}$ (Kaul, 1988 \cite{Kaul88});
{\bf e.}:
${\rm Fe_{16}Ni_{64}B_{19}Si_1}$ (Kaul {\em et al.}, 1994 \cite{Kaul94});
{\bf f.}:
${\rm Fe_{86}Co_{4}Zr_{10}}$ (Babu {\em et al.}, 1997 \cite{Babu97});
{\bf g.}:
${\rm Fe_{90}Zr_{10}}$ (Babu {\em et al.}, 1997 \cite{Babu97});
{\bf h.}:
${\rm Fe_{90-x}Mn_xZr_{10}}$ (Perumal {\em et al.}, 2001 \cite{Perumal01}).
}
\end{figure}

As far as the disorder should be irrelevant for the asymptotic
critical behaviour of the Heisenberg magnets, the diluted $d=3$ Heisenberg
magnets should belong to the same $O(3)$ universality class as the pure ones.
Theoretically predicted values of the isothermal magnetic susceptibility,
correlation length, heat capacity, pair correlation function, and the order
parameter {\em asymptotic} critical exponents in this universality class
read \cite{Guida98}:
\begin{eqnarray}\label{2}
&&
\gamma=1.3895\pm0.0050,\,\nu=0.7073\pm0.0035, \, \alpha=-0.122\pm0.009,\\
\nonumber &&
\eta=0.0355\pm0.0025,
\,\beta=0.3662\pm0.0025.
\end{eqnarray}
The experimental picture is more controversial. The bulk of experiments
on critical behaviour of disordered Heisenberg-like magnets performed
up to middle 80-ies is discussed in the comprehensive reviews
\cite{Egami84,Kaul85}.
More recent experiments may be found in
\cite{heis_all,Fahnle83,Kaul88,Kaul94,Babu97,Perumal01}
and references therein. We show typical results of measurements of
the isothermal magnetic susceptibility effective critical exponent
$\gamma_{\rm eff}$ (\ref{1}) in Figs. \ref{fig1}.
As it is seen from the pictures, the behaviour of $\gamma_{\rm eff}$ is
non-monotonic. The exponent differs from its value predicted in the asymptotic
limit (\ref{2}) and is a subject of a wide crossover behaviour. Before
reaching asymptotics  $\gamma_{\rm eff}$ possess maximum (except of the fig.
\ref{fig1}.d), the value of the maximum is system dependent: it
differs for different magnets.

It is standard now to rely on the renormalization group  (RG) method
\cite{rgbooks} to get a reliable quantitative description of the behaviour
in the vicinity of critical point. Namely in this way the cited above
values (\ref{2}) of the critical exponents of $d=3$
Heisenberg model were obtained. The RG approach appeared
to be a powerful tool to describe asymptotic \cite{review} and
effective \cite{Folk00} critical behaviour of disordered
Ising-like magnets as well. The purpose of the present paper is to
describe the crossover behaviour of disordered Heisenberg-like
magnets in frames of the field-theoretical RG technique. In
particular we want to calculate theoretically the isothermal magnetic
susceptibility effective critical exponent and to explain in this
way the appearance of the peak in its typical experimental
dependencies. The rest of the paper is organized as follows. In
the Section \ref{II} we formulate the model and review main
theoretical results obtained for it so far by means of the RG
technique, effective critical behaviour is analyzed in the Section
\ref{III}, we end by conclusions and outlook in the Section \ref{IV}.

%%%%%%%%%%%%%%%%%%%%%%%%%%%%%%%%%%%%%%%%%%%%%%%%%%%
%SECTION II
%%%%%%%%%%%%%%%%%%%%%%%%%%%%%%%%%%%%%%%%%%%%%%%%%%%
\section{The model and its RG analysis}
\label{II}

The model of a random quenched magnet we are going
to consider is described by the following Hamiltonian:
\begin{equation} \label{3}
H=-\frac{1}{2}\sum_{{\bf R},{\bf R'}}J(|{\bf R}-{\bf R'}|)
\vec{S}_{{\bf R}} \vec{S}_{{\bf R'}} c_{{\bf R}} c_{{\bf R'}}.
\end{equation}
Here, the sum spans over all sites ${\bf R}$ of $d$-dimensional
hypercubic lattice, $J(|{\bf R}-{\bf R'}|)$ is a short-range
(ferro)magnetic interaction between classical ``spins" $\vec{S}_{{\bf
R}}$ and  $\vec{S}_{{\bf R'}}$.  We consider the spins $\vec{S}_{{\bf
R}}$ to be $m$-component vectors and the Hamiltonian (\ref{3})
contains  their scalar product. Obviously, for the particular case of
Heisenberg spins we will put later $m=3$.  The randomness is
introduced into the Hamiltonian (\ref{3}) by the occupation numbers
$c_{\bf R}$ which are equal 1 if the site ${\bf R}$ is occupied by a
spin and $0$ if the site is empty. Considering the case when occupied
sites are distributed without any correlation and fixed in certain
configuration one obtains so-called  uncorrelated quenched $m$-vector
model.

In principle, the above information is enough to apply the RG approach
for a  study of the critical behaviour of the model (\ref{3}). One should
obtain an effective Hamiltonian corresponding to the model under
consideration and then one analyzes its long-distance properties by
analyzing appropriate RG equations \cite{rgbooks}. But already on this
step there are at least two different possibilities to proceed and
both were exploited for the model (\ref{3}). On one hand, to get the
free energy of the model one can average the logarithm of
configuration-dependent partition function over different possible
configurations of disorder \cite{Brout59}. Then, making use of the
replica trick \cite{Emery75} one arrives to the familiar effective
Hamiltonian \cite{Grinstein76}:
\begin{equation}\label{4}
H_{\rm eff} {=} {-}\!\int\!\! d^dR\!
\left\{\!\frac{1}{2}\sum_{\alpha{=}1}^n\left[{\mu_0}^2|\vec{\phi}^\alpha|^2{+}
|\vec{\nabla}\vec{\phi}^\alpha|^2\right]\!{+}
\frac{u_0}{4!}\sum_{\alpha{=}1}^n|\vec{\phi}^\alpha|^4{+}
\frac{v_0}{4!}\!{\left(\sum_{\alpha{=}1}^n |\vec{\phi}^\alpha|^2\right)\!}^2
\!\right\}
\end{equation}
describing in the replica limit $n\rightarrow 0$
critical properties of the model (\ref{3}).
Here, $\mu_0$ is a bare mass, $u_0>0$  and $v_0\leq0$ are
bare couplings and $\vec{\phi}^{\alpha} \equiv \vec{\phi}^{\alpha}({\bf R})$
is an $\alpha$-replica of
$m$-component vector field. The prevailing amount of RG studies of
the critical behaviour of  quenched $m$-vector model was performed on the
base of the effective Hamiltonian (\ref{4}) \cite{review}.

However, one more effective Hamiltonian corresponding to the model
(\ref{3}) is discussed  in the literature
\cite{Sobotta78,Sobotta80,Sobotta82,Sobotta85}.
It is obtained exploiting the idea that a quenched disordered system
can be described as an equilibrium system with additional forces of
constraints  \cite{Morita64}. In such approach both variables
$\vec{S}_{\bf R}$ and $c_{\bf R}$ are treated equivalently and one
ends up with the effective Hamiltonian which differs from (\ref{4})
and, consequently, leads to different results for the critical
behaviour of the model (\ref{3})
\cite{Sobotta78,Sobotta80,Sobotta82,Sobotta85}.
Whereas the effective Hamiltonian (\ref{4}) was used in the wide
context of general $m$-vector models \cite{review},
the approach of Refs.
\cite{Sobotta78,Sobotta80,Sobotta82,Sobotta85} was mainly used in
explanations of crossover behaviour in Heisenberg-like systems
\cite{west}. Below, we will discuss our results, based on
the effective Hamiltonian (\ref{4})  for $m=3$ and compare them
with those derived in
\cite{Sobotta78,Sobotta80,Sobotta82,Sobotta85}.

As it is well known, the renormalization group (RG) approach makes use of
the scaling symmetry of the system in the asymptotic limit to extract the
universal content and at the same time removes divergencies which occur for
the evaluation of the bare functions in this limit \cite{rgbooks}.
A change in the renormalized couplings $u$, $v$ of the effective Hamiltonian
(\ref{3}) under the RG transformation is described by the flow equations:
\begin{equation} \label{5}
\ell\frac{\rm d}{{\rm d} \ell}u(\ell)=\beta_u\left(u(\ell),v(\ell)\right),\quad
\ell\frac{\rm d}{{\rm d}\ell}v(\ell)=\beta_v\left(u(\ell),v(\ell)\right).
\end{equation}
Here, $\ell$ is the flow parameter related to the distance $\tau$ to the critical
point. The fixed points ($u^*,v^*$) of the system of differential equations
(\ref{5}) are given by:
\begin{equation} \label{6}
\beta_u\left(u^*,v^*\right)=0,\quad
\beta_v\left(u^*,v^*\right)=0.
\end{equation}
A fixed point is said to be stable if the stability matrix
\begin{equation}\label{6a}
B_{ij}\equiv\partial \beta_{u_i}/\partial u_j, \hspace{3em}
i,j=1,2;
\hspace{3em}
u_i=\{u,v\},
\end{equation}
possess in this point eigenvalues $\omega_1,\omega_2$ with positive real parts.
In the limit $\ell\to 0$, $u(\ell)$ and $v(\ell)$ attain the
stable fixed point values $u^*,v^*$. If the stable fixed point
is reachable from the initial conditions (let us recall that for the
effective Hamiltonian (\ref{3}) they read $u> 0,v \leq 0$) it corresponds to
the critical point of the system. The asymptotic critical exponents values
are defined by the fixed point values of the RG
$\gamma$-functions. In particular the isothermal magnetic susceptibility
exponent $\gamma$ is expressed in terms of the RG functions $\gamma_{\phi}$
and ${\bar \gamma}_{\phi^2}$ describing renormalization of the field $\phi$
and of the two-point vertex function with a $\phi^2$ insertion
correspondingly \cite{rgbooks}:
\begin{equation} \label{7}
\gamma^{-1}=1 - \frac{{\bar \gamma}_{\phi^2}}{2-{\gamma}_{\phi}}.
\end{equation}
In Eq. (\ref{7}), the functions  $\gamma_{\phi}\equiv \gamma_{\phi}(u,v)$,
${\bar \gamma}_{\phi^2}\equiv{\bar
\gamma}_{\phi^2}(u,v)$ are calculated in  the stable fixed
point $u^*,v^*$. In the RG scheme, the effective critical exponents
are calculated in the region,
where couplings $u(\ell),v(\ell)$ have not reached their fixed point values
and depend on $\ell$. In particular for the exponent $\gamma_{\rm eff}$ one
gets: \begin{equation}\label{8}
\gamma^{-1}_{\rm eff}(\tau)=1 - \frac{{\bar \gamma}_{\phi^2}
[u\{\ell(\tau)\},v\{\ell(\tau)\}]}{2-{\gamma}_{\phi}
[u\{\ell(\tau)\},v\{\ell(\tau)\}]}+\dots.
\end{equation}
In (\ref{8}) the  part denoted by dots
is proportional to the $\beta$--functions
(\ref{5}) and comes from the change of the amplitude part of the susceptibility.
In the subsequent calculations we will neglect this part, taking
the contribution of the amplitude function to the crossover to be small
\cite{note2}.

\begin{figure}[htbp]
\begin{center}
{\includegraphics%[width=0.7\textwidth]
{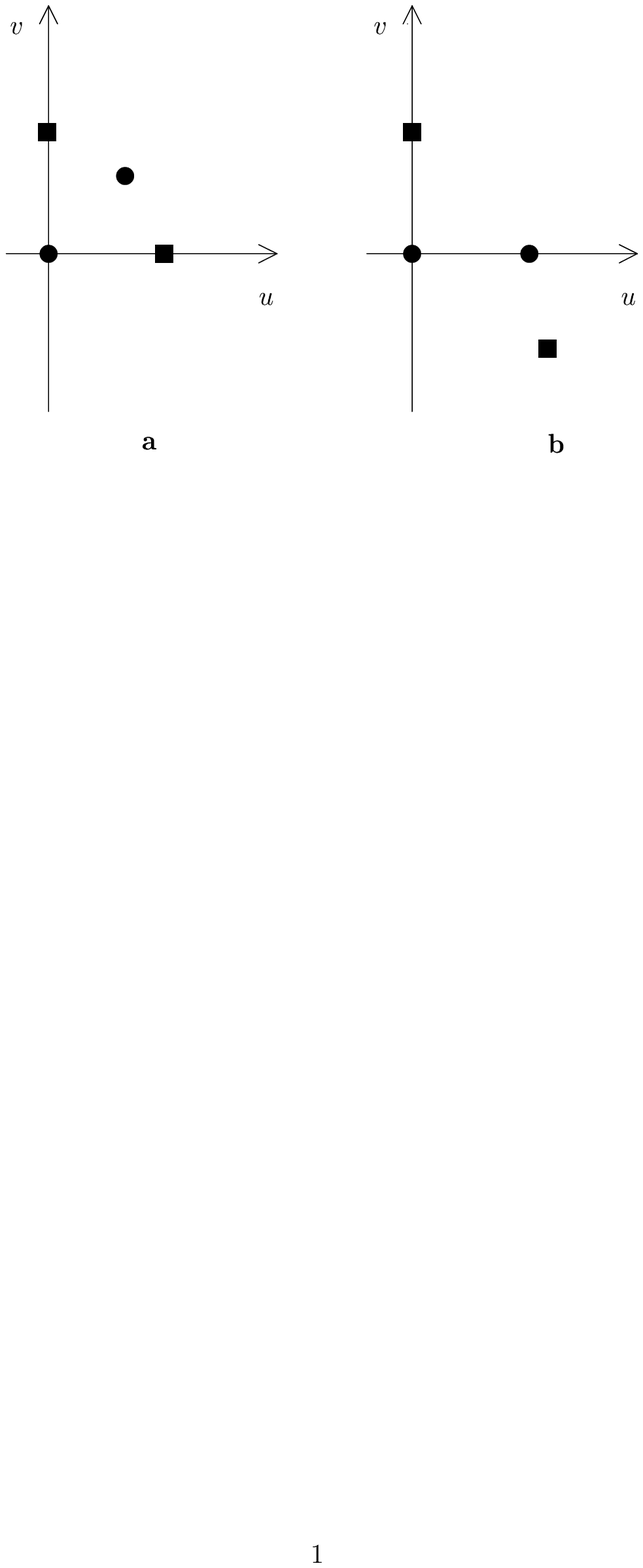}}
\end{center}
\caption{\label{fig2} Fixed points structure for the effective
Hamiltonian (\ref{4}) at $d=3$ and arbitrary $m$. {\bf a}:
$m>m_c$, {\bf b}: $m<m_c$. Stable fixed points are shown by filled
boxes, unstable ones are shown by filled circles. Only stable
fixed points with coordinates $u^*> 0$, $v^*\leq0$ are reachable
for the model of the quenched magnet (\ref{3}). }
\end{figure}

For the effective Hamiltonian (\ref{4}), the fixed point structure is
well established \cite{review}.
It is schematically shown in Figs. \ref{fig2}.a, \ref{fig2}.b.
Two qualitatively  different scenarios are observed:
for $m>m_c$ the critical behaviour of the disordered magnet is
governed by the fixed point of the pure magnet ($u^*>0$, $v^*=0$),
whereas for $m<m_c$ the new stable fixed point ($u^*>0$, $v^*<0$)
governs the asymptotic critical behaviour of the disordered magnet.
At the marginal dimensionality $m_c$ which separates these two
regimes, the $\alpha$ exponent of the pure magnet equals zero in
agreement with the Harris criterion.

Best theoretical estimates of
$m_c$ definitely support $m_c<2$: $m_c=1.942\pm0.026$
\cite{Bervillier86}, $m_c=1.912\pm0.004$ \cite{Dudka01}. Consequently,
the fixed point structure of the model of diluted Heisenberg-like magnet
($m=3$) is given by Fig. \ref{fig2}.{\bf a}: the stable reachable
fixed points of the diluted and pure Heisenberg-like magnets do coincide
($u^*\neq 0,v^*=0$),
hence their {\em asymptotic} critical exponents do coincide as well.
However the last statement does not concern the {\em effective} exponents.
These are defined by the running values of the couplings
$u(\ell)\neq 0,v(\ell)\neq 0$ and will be calculated in the next
section.

%%%%%%%%%%%%%%%%%%%%%%%%%%%%%%%%%%%%%%%%%%%%%%%%%%%
%SECTION III
%%%%%%%%%%%%%%%%%%%%%%%%%%%%%%%%%%%%%%%%%%%%%%%%%%%
\section{The RG flows and the effective critical behaviour}
\label{III}

The RG functions of the model (\ref{4}) are known by now in
pretty high orders of the perturbation theory
\cite{review,Pelissetto00}.
For the purpose of present study we will restrict ourselves by the
first approximation where the described crossover phenomena
manifests itself for the Heisenberg-like disordered magnets in
non-trivial way. Within the two loop approximation in the
minimal subtraction RG scheme \cite{Hooft72} the RG-functions read
\cite{Kyriakidis96}:
\begin{eqnarray}
 \nonumber
\beta_u(u,v)&=&-u(\varepsilon-\frac{m+8}{6}{u}-2v+
\frac{3m+14}{12}u^2+{\frac {5mn+82}{36}}v^2+\\ \label{9}
&&
\frac{11m+58}{18}uv),
\\ \nonumber
\beta_v(u,v)&=&-v(\varepsilon-\frac{m+2}{3}u-\frac{mn+8}{6}v+
\frac{5(m+2)}{36}u^2 + \\ \label{10}
&&
\frac{3mn+14}{12}v^2+
\frac{11(m+2)}{18}uv),
\\ \label{11}
\gamma_{\phi}(u,v)&=&\frac{m+2}{72}u^2+\frac{mn+2}{72}v^2+
\frac{m+2}{36}uv,
\\ \label{12}
\bar{\gamma}_{\phi^2}(u,v)&=&\frac{m+2}{6}u+\frac{mn+2}{6}v
-\frac{m+2}{12}{u}^{2}-\frac{mn+2}{12}v^2-\frac{m+2}{6}uv.
\end{eqnarray}
Here, $\varepsilon=4-d$ and replica limit $n=0$ is to be taken.

Starting form the expressions (\ref{9})--(\ref{12}) one can
either develop the $\varepsilon$-expansion, or work directly at
$d=3$ putting in (\ref{9}), (\ref{10}) $\varepsilon=1$ and
considering renormalized couplings $u,v$ as the expansion
parameters \cite{Schloms}.
However, such RG perturbation theory series with several couplings
are known to be asymptotic at best \cite{rgbooks}.
One should apply appropriate resummation technique to improve their
convergence to get reliable numerical data on their basis.
We used several different resummation schemes for this purpose.
Here we will give the results obtained by the method which allowed
to analyze the largest region in the parametric $u-v$ space.
The method was proposed in Ref. \cite{Alvarez00} and was successfully
applied to study random $d=3$ Ising model  \cite{Pelissetto00}.
Moreover, it was shown that the RG functions of the $d=0$ random Ising
model are Borel-summable by this method \cite{Alvarez00}.
The main idea proposed in Ref. \cite{Alvarez00} is to consider
resummation in variables $u$ and $v$ separately. Taken that the RG
function $f(u,v)$ is given to the order of $p$ loops, one first rewrites
it as a power series in $v$:
\begin{equation} \label{13}
f(u,v)=\sum_{k=0}^pA_k(u)v^k.
\end{equation}
Then each coefficient $A_k(u)$ is considered as power series in $u$ and
resum\-med as a function of a single variable $u$ thus obtaining the
resummed functions $A^{res}_k(u)$. Next one substitutes these
functions into (\ref{13}) and resums  the RG function $f$ in single
variable $v$. For the resummation in a single variable one may use
any of familiar methods. Our results are obtained by making use of
the  Pad\'e-Borel-Leroy method \cite{Nickel78}.

First, applying the above described resummation procedure to the
$\beta$-functions (\ref{9}), (\ref{10}) we get the pure Heisenberg
fixed point coordinates $u^*=0.8956$, $v^*=0$. The stability matrix
(\ref{6a}) eigenvalues are
positive at this fixed point ($\omega_1=0.577$,
$\omega_2=0.147$) providing its stability. Then for the resummed
values of the asymptotic critical exponents we get \cite{note3}:
\begin{equation}\label{14}
\gamma=1.382,\,\nu=0.701, \, \alpha=-0.104,\,
\eta=0.030,
\,\beta=0.361.
\end{equation}
We do not give the confidence intervals in (\ref{14}), as far as
they can be estimated only by comparison of changes introduced by
different orders of perturbation theory. Note however that the
results (\ref{14}) are in a good agreement with the most accurate
estimates of the exponents in the $O(3)$ universality class (\ref{2}).
This brings about that both the considered here two-loop
approximation as well as the chosen resummation technique give an
adequate description of  asymptotic critical phenomena.

Before passing to the effective critical exponents let us first
analyze the corrections to scaling.
For the pure Heisenberg magnet, taking into account the
leading correction to scaling results in the following
formula for the
isothermal susceptibility:
\begin{equation}\label{15}
\chi(\tau)=\Gamma_0\tau^{-\gamma}(1+\Gamma_1\tau^{\Delta}),
\end{equation}
where the correction-to-scaling exponent
is given by $\Delta=\omega\nu$ with
$\omega=\partial\beta_u(u)/\partial u
|_{u=u^*}$  and non-universal critical amplitudes $\Gamma_0, \Gamma_1$.
For the diluted  Heisenberg  magnet the corresponding formula
includes two leading corrections $\Delta_1$, $\Delta_2$
(see e.g. \cite{Kaul88}):
\begin{equation}\label{16}
\chi_(\tau)=\Gamma'_0\tau^{-\gamma}(1+\Gamma'_1\tau^{\Delta_1}+
\Gamma'_2\tau^{\Delta_2}),
\end{equation}
with critical amplitudes  $\Gamma'_0, \Gamma'_1,\Gamma'_2$.
The exponents $\Delta_i$ are expressed in terms of the
stability matrix (\ref{6a}) eigenvalues $\omega_i$ in the pure
Heisenberg fixed point: $\Delta_i=\nu \omega_i$.
At this fixed point, the eigenvalues of the stability matrix
(\ref{6a}) read:
\begin{equation}\label{17}
\omega_1=\frac{\partial\beta_u(u,v)}{\partial u}|_{u^*\neq0,v^*=0},
\hspace{3em}
\omega_2=\frac{\partial \beta_v(u,v)}{\partial
v}|_{u^*\neq0,v^*=0}.
\end{equation}
It is straightforward to see that the value $\omega_1$ coincides with
the exponent $\omega$ of the pure model whereas it may be shown
(see e.g. \cite{Kaul88,Kyriakidis96}) that
the exponent $\omega_2=|\alpha|/\nu$ where $\alpha$ and $\nu$ are
the heat capacity and correlation length critical exponent of the pure
Heisenberg model. On the base of the numerical values of the exponents
(\ref{14}) we get:
\begin{equation}\label{18}
\Delta_1=0.405, \hspace{3em}
\Delta_2=0.104.
\end{equation}
Again, obtained by us in the two-loop approximation numbers
(\ref{18}) can be compared with those in the six-loop
approximation making use of the data (\ref{2}) together with the
value of $\omega$ of pure 3d Heisenberg model
$\omega=0.782\pm0.0013$
\cite{Guida98}. As we have noted above, in order to get the
numerical values of the correction-to-scaling exponents of
diluted Heisenberg model it is no need to consider the RG
functions (\ref{9})--(\ref{12}) in the whole region of couplings
$u,v$: it is enough to know them for the case of the pure model
(i.e. for $u\neq0,v=0$). However, to get the effective exponents
it is necessary to study complete set of the RG functions
(\ref{9})--(\ref{12}) working also in the region where both
couplings $u$ and $v$ differ from zero.

\begin{figure}[htbp]
{\includegraphics[width=0.7\textwidth] {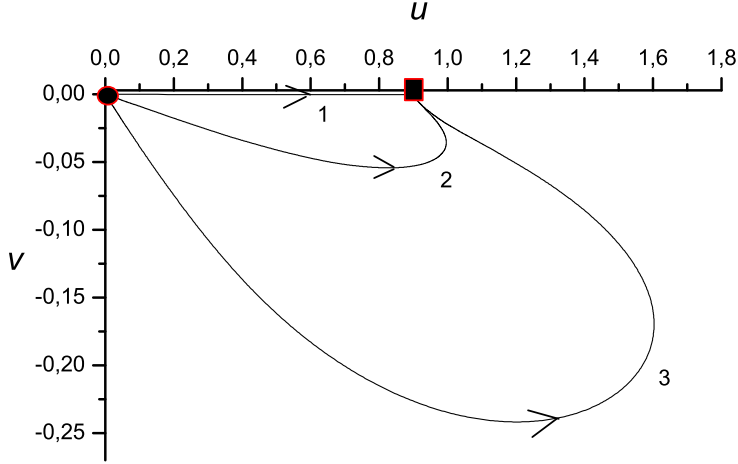}} \caption{
\label{fig3} Flows in the parametric space of couplings. The
filled box denotes the stable fixed point $u^*=0.8956,\,v^*=0$.
Curve 1 corresponds to the flow from initial values with $v_0=0$,
curve 2 starts with a small ratio $|v_0/u_0|$ whereas flow 3
corresponds to larger $|v_0/u_0|$.}
\end{figure}

To this end we use the above described
resummation technique in order to restore the convergence of the
RG expansions in couplings $u$, $v$. First we solve the system of
differential equations (\ref{5}) and get the running values of
couplings $u(\ell)$, $v(\ell)$ (\ref{9})--(\ref{12}).
They define the flow in the parametric
space $u,v$ and in the limit $\ell\rightarrow 0$ attain the stable
fixed point value (shown by the filled box in Fig. \ref{fig3}).
Character of the flow depends on the initial conditions $u_0,v_0$
for solving the system of differential equations (\ref{5}).
For the model (\ref{3}), the coupling $v$ is
proportional to variance of disorder \cite{review}
thus one can use the ratio $|v_0/u_0|$ to define the degree of dilution.
Typical flows which are obtained for different ratios $|v_0/u_0|$
are shown in Fig. \ref{fig3} by curves 1-3. We choose the starting
values in the region with the appropriate signs of couplings
$u>0,\,v<0$ near the origin (in the vicinity of the Gaussian fixed point
$u^*=v^*=0$ shown by the filled circle in the figure). The flow No
1 is obtained for $v_0=0$, it corresponds to the pure Heisenberg
model. The flow No 2 results from the small ratio $|v_0/u_0|$ and
corresponds to the weak disorder whereas the flow No 3 is obtained
for large $|v_0/u_0|$ and corresponds to the stronger dilution.

\begin{figure}[htbp]
{\includegraphics[width=0.7\textwidth] {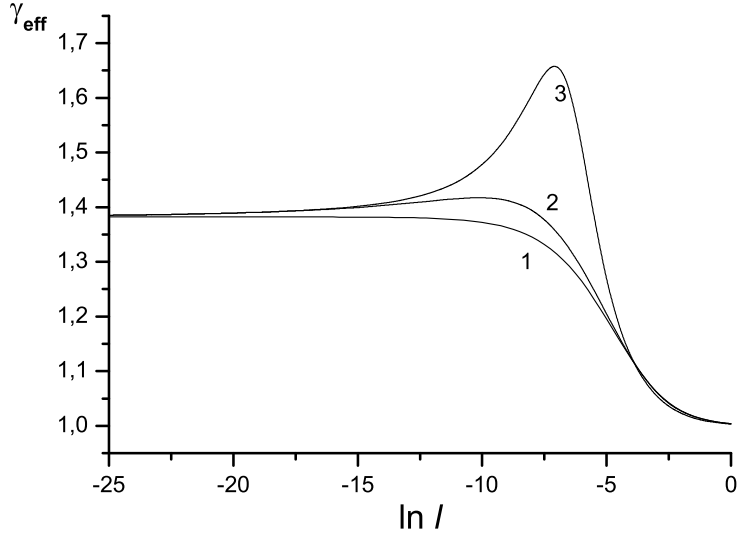}} \caption{
\label{fig4} Effective critical exponent $\gamma_{\rm eff}$ versus
logarithm of the flow parameter.  The curves correspond to flows
from the Fig. \ref{fig3} denoted by corresponding numbers.}
\end{figure}

Obtained running values of coupling constant presented by flows in
Fig. \ref{fig3} allow one to get the effective
critical exponents. Calculating resummed expression for the effective
exponent $\gamma_{\rm eff}$ (\ref{8}) along the flows 1-3 we get the
results shown in the Fig. \ref{fig4}.
Again, the curve 1 corresponds to the effective critical exponent
of the pure Heisenberg model, whereas curves 2 and 3 provide two
possible scenarios for the effective exponents of the disordered
Heisenberg model. Curve 2 corresponds to the weak dilution region:
here, the exponent increases with approach to the
critical point, although the crossover region is larger in
comparison with the pure magnet (compare curves 1 and 2 in Fig.
\ref{fig4}). This may lead to the peculiar situation that the
asymptotic value of the exponent is reached earlier than the
asymptotic values of the coupling.
The effective exponents for the flows
originating from non-zero ratio $|v_0/u_0|$ always attain the value
which are larger than the asymptotic one. But the absolute value of this
``overshooting" for small enough $|v_0/u_0|$ is too small to be
observed experimentally.
An experimental observation of such type of
$\gamma_{\rm eff}$  behaviour of the disordered Heisenberg-like magnet
is provided e.g. by Fig. \ref{fig1}{\bf d}. Different behaviour of
$\gamma_{\rm eff}$ is demonstrated by the curve 3 in Fig. \ref{fig4}.
Here, before reaching the asymptotic region the exponent
possess a distinct peak. Such behaviour is in
agreement with observed experimental data presented by Figs.
\ref{fig1}{\bf a}--\ref{fig1}{\bf c},
\ref{fig1}{\bf e}--\ref{fig1}{\bf h}. The value of
maximum depends on the initial values
for the RG flows. Larger ratio $|v_0/u_0|$  (i.e. stronger disorder)
leads to the larger maximum.
Thus, within unique approach one may explain both scenarios
observed in the diluted Heisenberg-like magnets effective critical exponent
$\gamma_{\rm eff}$ behaviour.

As we have noticed in the section \ref{II}, the crossover behaviour of
random Heisenberg-like magnets was analyzed by means of an alternative
approach in \cite{Sobotta78,Sobotta80,Sobotta82,Sobotta85}. There, the
quenched disordered magnet was described as an equilibrium
one with additional forces of constraints  \cite{Morita64}.
This resulted  in an effective Hamiltonian which differs from
(\ref{4}). The fixed point structure of this Hamiltonian differs from
those given in Fig. \ref{fig2}  and, for different concentrations,
leads to different crossover regimes. In particular, it predicts that
there exists a limiting value of concentration where the critical
behaviour is governed by Fisher-renormalized tricritical exponents
\cite{Sobotta85} which coincide with those of a $d=3$ spherical
model: $\gamma=2$, $\nu=1$, $\alpha=-1$, $\eta=0$, $\beta=1/2$.
There exist two more fixed points which may be stable in the weak
dilution regime. Their stability differs in different orders of
the perturbation theory (compare \cite{Sobotta78} and
\cite{Sobotta82}) but the numerical values of the critical
exponents do not differ essentially at these fixed points. The
maximal possible value of the effective critical exponent
$\gamma_{\rm eff}$ has been estimated as $\gamma_{\rm eff}\simeq 2.6$
\cite{Sobotta82}. However, the distinct
feature of the behaviour of $\gamma_{\rm eff}(\tau)$ obtained in
\cite{Sobotta80} is its monotonic
dependence. Hence, the experimentally observed peaks (see Fig.
\ref{fig1}) can not be explained within such approach.

%%%%%%%%%%%%%%%%%%%%%%%%%%%%%%%%%%%%%%%%%%%%%%%%%%%
%SECTION IV
%%%%%%%%%%%%%%%%%%%%%%%%%%%%%%%%%%%%%%%%%%%%%%%%%%%
\section{Conclusions}
\label{IV}

In the present paper we used the field-theoretical RG technique to
study the effective critical behaviour of diluted Heisenberg-like
magnets. The question of particular interest was to explain the
peak in the exponent $\gamma_{\rm eff}$ as function of distance from
$T_c$ observed in some experiments.
Our two-loop calculations refined by
the resummation of the perturbation theory series resulted in
typical behaviour of diluted Heisenberg-like magnets
$\gamma_{\rm eff}$ exponent represented by curves 2 and 3 in Fig.
\ref{fig4}. The exponent can either reach it asymptotic value
without demonstrating distinct maximum or it can first reach the
peak and then cross-over to the asymptotic value from above. The
strength of disorder is a physical reason which discriminates
between these two regimes.

Our calculations are quite general and do not specify any particular
object. In order to fit our curves to certain experiment one should
include into consideration non-universal parameters to specify the
magnetic system. The same concerns the flow parameter $\ell$ which
as we have already noted is related to the distance to the
critical point $\tau$. In principle such calculations may be done.
However we want to emphasize that our analysis shows the reason of
the peak in $\gamma_{\rm eff}(\tau)$ dependence for different disordered
magnets which belong to the $O(3)$ universality class and this reason may
be explained within the traditional RG approach. This concerns not only
the magnetic susceptibility effective critical exponent.
One more example is given by the order parameter effective exponents
$\beta_{\rm eff}$ which has minimum when $\tau$  goes to zero
(see e.g. \cite{Perumal01}). Interpretation of this effect will be the
goal of a separate study

In conclusion we want to note that similar peculiarities of
the effective critical behaviour may be
observed in studies of disordered easy-plane magnets which belong
to the $O(2)$ universality class. Since the heat capacity does not
diverge in such systems, the RG fixed point scenario is given by the
Fig. \ref{fig2}{\bf a} as for the Heisenberg-like disordered magnets.
Up to our knowledge such experiments have not been performed yet and
we hope that our calculations may stimulate them.

M. D. acknowledges the Ernst Mach research fellowship of the
\"Osterreichisher Austauschdienst. This work
was supported in part by \"Osterreichische Nationalbank
Jubil\-\"aums\-fonds through grant No 7694.


\begin{thebibliography}{90}
\bibitem{note1} In experimental studies, random Ising-like magnets can
be obtained as a crystalline mixture of two compounds on the basis of
``pure'' Ising model matrix. For example, corresponding site-diluted
uniaxial alloy can be prepared by substitution of ${\rm FeF_2}$
(${\rm MnF_2}$) with its non-magnetic isomorph ${\rm ZnF_2}$
\cite{review}.
\bibitem{note3} We calculate exponents $\gamma$ and $\nu$ from the
relations (\ref{7}) and
$\nu^{-1}=2-\bar{\gamma}_{\phi^2}-\gamma_{\phi}$.
Numerical values of the rest of
exponents are obtained from the familiar scaling relations.
\end{thebibliography}
\end{document}